\documentclass[aps,prd,onecolumn,11pt,groupedaddress,nofootinbib,showpacs]{revtex4}
\usepackage{amsmath}
\begin{document}
\title{
The Ising spin glass in finite dimensions: a perturbative study of the
free energy}

\author{T.~Temesv\'ari}
\email{temtam@helios.elte.hu}
\affiliation{
Research Group for Theoretical Physics of the Hungarian Academy of Sciences,
E\"otv\"os University, P\'azm\'any P\'eter s\'et\'any 1/A,
H-1117 Budapest, Hungary}

\date{\today}

\begin{abstract}
Replica field theory is used to study the $n$-dependent free energy of the
Ising spin glass in a first order perturbative treatment.
Large sample-to-sample deviations of the free energy from its quenched average
prove to be Gaussian, independently of the special structure of the order
parameter. The free energy difference between the replica symmetric and
(infinite level) replica symmetry broken phases is studied in details:
the line $n(T)$ where it is zero coincides with the Almeida-Thouless
line for $d>8$. The dimensional domain $6<d<8$ is more complicated, and 
several scenarios are possible.
\end{abstract}

\pacs{75.10.Nr}

\maketitle

\section{Introduction}

Large scale free energy deviations $F\equiv Nf$ of a quenched disordered
system from its selfaveraged mean $Nf_{\text{typ}}$ are extremely
rare events with probability $\sim e^{N\,L(f)}$, with the system size $N
\to\infty$, and $L(f)=O(1)<0$ for $f=O(1)\not =f_{\text{typ}}$. Only the
case $f<f_{\text{typ}}$ is considered in this paper. As invented in
Ref.\ \cite{Crisanti_et_al}, $L(f)$ can be computed by the replica method;
namely it is the Legendre transform of the $n$-dependent free energy $\Phi$
of (\ref{Phi}):
\[
\beta^{-1}L(f)=-n\Phi+nf\qquad\text{and}\qquad \frac{d}{dn}(n\Phi)=f.
\]
The replica number $n$ is considered a nonnegative real number corresponding
to $f\le f_{\text{typ}}$, see also \cite{Parisi_Rizzo_1,Parisi_Rizzo_2}.
In most applications, we have $\Phi-f_{\text{typ}}\sim -n^{a}$ for $n\to
0^{+}$ with some positive integer exponent $a$ characteristic to the
model, and also to the region of control parameters (temperature, magnetic
field, etc.) considered. Legendre transforming then provides $L(f)\sim
-(f_{\text{typ}}-f)^{\frac{a+1}{a}}$. In the most common situations
\cite{Crisanti_et_al,Parisi_Rizzo_1,Parisi_Rizzo_2}, $a=1$ and the large
scale free energy fluctuations are Gaussian.

A very atypical behaviour was found many years ago by Kondor \cite{Kondor}
for the truncated version of the Sherrington-Kirkpatrick (SK) model \cite{SK},
i.e.\ for the mean field Ising spin glass, just below the transition
temperature. The exponent $a=5$ was later proved to be true in the whole
spin glass phase, including zero temperature
\cite{Parisi_Rizzo_1,Parisi_Rizzo_2}. The original motivation of the present
work was to find out how the large free energy deviations are influenced
by including the geometry of a $d$-dimensional hypercubic lattice. For that
purpose, we followed the usual steps to generate a field theory convinient
for a perturbative calculation whose zeroth order is the mean field (SK)
theory. These steps are:
\begin{itemize}
\item Transformation of the lattice system to a field theory, see
Refs.\ \cite{BrMo79,rscikk}. The natural ultraviolet cutoff is then the
boundary of the Brillouin zone, and it is $\Lambda=O(1)$ when length is
measured in the units of the lattice spacing.
\item The momentum space is shrinking such that $T_c-T\sim \tau\ll
\Lambda\ll 1$.
\item Momentum dependence is truncated up to the quadratic term.
\end{itemize}
We are led then to the replicated field theory presented in the next
section, Eqs.\ (\ref{Z^n}) and (\ref{L}).
(Zero external magnetic field is assumed throughout the present paper.)
It turns out from the calculation of Sec.\ \ref{IV}
that the leading behaviour of the large scale fluctuations changes to
Gaussian when first order corrections to the low temperature SK model are
taken into account (this has been noticed in Ref.\ \cite{As_Mr}), and the
anomalous $n^5$ term is now subleading. It becomes also clear that this
Gaussian behaviour follows for any structure of the order parameter as long
as replica equivalence is assumed.

The glassy phase of the SK model has the ultrametric structure with infinite
step replica symmetry breaking \cite{MePaVi} (RSB)\footnote{The acronym
RSB is used throughout the paper for the special type of replica symmetry
breaking with infinite number of levels in the ultrametric hierarchy.}.
For $T<T_c$, this RSB state is reached by a phase transition \cite{Kondor}
when $n$ is lowered from around $1$ --- where $\Phi(n)$ is the free energy of
the stable replica symmetric (RS) phase with a nonzero order parameter
--- at some $n_{\text{AT}}(T)$ having two characteristics:
\begin{enumerate}
\item The RS phase becomes unstable due to the Almeida-Thouless \cite{AT}
instability, i.e.\ the so called replicon mass $\Gamma_\text{R}$ is zero for
$n=n_{\text{AT}}(T)$.
\item $\Phi_{\text{RS}}=\Phi_{\text{RSB}}$ at the transition.
\end{enumerate}
Whether or not this feature of the SK model is perturbatively stable is
studied in the second part of the paper, and rather nontrivial calculations
lead us to a positive conclusion 
for $d>8$. The transitional domain of dimensions
above the upper critical dimension ($6<d<8$) is, however, more complicated:
it is argued that the question cannot be settled by
this first order calculation.

The outline of the paper is as follows: Sec.\ \ref{II} presents the replicated
field theoretical model suitable for the perturbative calculation, whose
basic formulae are provided in Sec.\ \ref{III}. The one-loop results for
the $n$-dependent free energy are shown in Sec.\ \ref{IV} up to fourth
order in the double series of $\tau\sim T_c-T$ and $n$: these are the leading
contributions when $d>8$. Nonanalytic temperature dependences are more and
more important while approaching dimension 8, and even for $6<d<8$. They are
computed for both the RSB and RS schemes in  Sec.\ \ref{V}. The line of
equal free energies of the two schemes is calculated for $d>8$ in
Sec.\ \ref{VI}, whereas we analyze the situation in the transitional
regime just above the upper critical dimension, i.e.\
$6<d<8$, in Sec.\ \ref{VII}. One-loop results for the RS replicon mass
$\Gamma_\text{R}$ are presented and the AT instability line
deduced in Sec.\ \ref{VIII}. Some concluding remarks are left to Sec.\ \ref{IX}.
Three appendices contain some computational details.

\section{Field theoretic model for the calculation of the free energy}
\label{II}

In this section, we will define the replicated field theory which is
appropriate for the calculation of perturbative corrections to the
$n$-dependent free energy $\Phi(n,\beta)$ defined as
\begin{equation}\label{Phi}
\beta \Phi(n,\beta)\equiv -\frac{1}{nN}\,\ln \overline{Z^n}
\end{equation}
where $Z$ is the partition function of an Ising spin glass defined on the
$d$-dimensional hypercubic lattice consisting of $N$ Ising spins, and the
bar denotes averaging over the quenched Gaussian disorder with zero mean
and variance $J^2$. The thermodynamic limit $N\to \infty$ is always understood,
whereas $n>0$ is small but kept finite. We follow the strategy suggested in
Ref.\ \cite{AT2008}: by conviniently choosing the bare parameters, the field
theory provides the SK results \cite{Kondor,Parisi_Rizzo_1,Parisi_Rizzo_2}
in the tree approximation, i.e.\ when
neglecting loops.
In zero external magnetic field we can express $\overline{Z^n}$ in the
following functional integral form (see Refs.\ \cite{rscikk,droplet}):
\begin{equation}\label{Z^n}
\overline{Z^n}=C\times \int[d\phi]\,e^{-\mathcal{L}},
\end{equation}
with the normalization factor
$C$ to be fixed later by choosing it to match the SK result.
The Lagrangian $\mathcal{L}$ has the form
\begin{multline}\label{L}
\mathcal{L}=
\frac{1}{2}\sum_{\mathbf p}
\bigg(\frac{1}{2} p^2+m\bigg)\sum_{\alpha\beta}
\phi^{\alpha\beta}_{\mathbf p}\phi^{\alpha\beta}_{-\mathbf p}
-\frac{1}{6\sqrt{N}}\,\sideset{}{'}\sum_{\mathbf {p_1p_2p_3}}
w\sum_{\alpha\beta\gamma}\phi^{\alpha\beta}_{\mathbf p_1}
\phi^{\beta\gamma}_{\mathbf p_2}\phi^{\gamma\alpha}_{\mathbf p_3}
-\frac{1}{24N}\,\sideset{}{'}\sum_{\mathbf {p_1p_2p_3p_4}}\\[2pt]
\bigg(u_1\!\!\sum_{\alpha\beta\gamma\delta}\phi^{\alpha\beta}_{\mathbf p_1}
\phi^{\beta\gamma}_{\mathbf p_2}\phi^{\gamma\delta}_{\mathbf p_3}
\phi^{\delta\alpha}_{\mathbf p_4}+u_2\!\sum_{\alpha\beta}
\phi^{\alpha\beta}_{\mathbf p_1}\phi^{\alpha\beta}_{\mathbf p_2}
\phi^{\alpha\beta}_{\mathbf p_3}\phi^{\alpha\beta}_{\mathbf p_4}
+u_3\!\sum_{\alpha\beta\gamma}\phi^{\alpha\gamma}_{\mathbf p_1}
\phi^{\alpha\gamma}_{\mathbf p_2}\phi^{\beta\gamma}_{\mathbf p_3}
\phi^{\beta\gamma}_{\mathbf p_4}+
u_4\!\!\sum_{\alpha\beta\gamma\delta}
\phi^{\alpha\beta}_{\mathbf p_1}\phi^{\alpha\beta}_{\mathbf p_2}
\phi^{\gamma\delta}_{\mathbf p_3}\phi^{\gamma\delta}_{\mathbf p_4}
\bigg)+\dots
\end{multline}
where the ellipses are for higher order replica symmetric invariants which
are consistent with the extra symmetry of the zero magnetic field subspace
\cite {droplet,nucl}.
In this 
$n(n-1)/2$ component field theory the fluctuating fields
are symmetric in the replica indices with zero diagonals:
$\phi^{\alpha\beta}_{\mathbf p}=\phi^{\beta\alpha}_{\mathbf p}$ and
$\phi^{\alpha\alpha}_{\mathbf p}=0$, $\alpha$,$\beta=1,\dots,n$.
(Momentum conservation is indicated by the primed
summations.) The bare mass $m$ depends on temperature as
$m=-\frac{k^2}{2J^2}({T_c^{\text{mf}}}^2-T^2)$ where $k$ is Boltzmann's
constant, and $kT_c^{\text{mf}}=J$ --- the mean field critical temperature
--- differs from the {\em exact\/} one, $T_c$. Introducing the new parameter
$\tau\equiv \frac{k^2}{2J^2}(T_c^2-T^2)$ measuring the distance from
criticality, the bare mass can be written as $m=m_c-\tau$ with $m_c$ being
obviously one-loop order. Beside $\tau$, the couplings $w$, $u_1$, $u_2$,
$u_3$, $u_4$, \dots parametrize the Lagrangian. Around the critical point
they can be considered as constants; a complete matching with the SK results
is achieved by choosing $w=1$, $u_1=3$, $u_2=2$, $u_3=-6$, $u_4=0$
\cite{AT2008}.

The spin glass phase below $T_c$ is characterized by the nonzero order
parameter $\phi^{\alpha\beta}\equiv \frac{1}{\sqrt{N}} \langle
\phi^{\alpha\beta}_{\mathbf p=0} \rangle$ where the average is now taken
by the measure proportional to $e^{-\mathcal L}$. It is useful to redefine
the fields by the shift $\phi^{\alpha\beta}_{\mathbf p} \longrightarrow
\phi^{\alpha\beta}_{\mathbf p}-\sqrt{N}\,\phi^{\alpha\beta}\,
\delta_{\mathbf p=0}^{\text{Kr}}$. By this transformation, the new fields
continue fluctuating around zero; on the other hand, however, the Lagrangian
has lost the higher symmetry of the 
paramagnetic phase resulting in the following generic theory (equally
convinient for an RS or RSB ansatz):\footnote{For the sake of simplifying
the notations, we will keep the symbols $\phi^{\alpha\beta}_{\mathbf p}$
and $\mathcal L$ for the transformed quantities.}
\[\mathcal L=\mathcal L^{(0)}+\mathcal L^{(1)}+\mathcal L^{(2)}+
\mathcal L^{(3)}+\mathcal L^{(4)}+\dots.
\]
The terms above can be worked out using the results of the Appendices B
and D of Ref.\ \cite{nucl} providing
\begin{align}
\frac{1}{N}\mathcal L^{(0)}&=\frac{1}{2}m\sum_{\alpha\beta}{\phi^{\alpha\beta}} ^2
-\frac{1}{6}w\sum_{\alpha\beta\gamma}\phi^{\alpha\beta}
\phi^{\beta\gamma}\phi^{\gamma\alpha}
-\frac{1}{24}\bigg(u_1\sum_{\alpha\beta\gamma\delta}\phi^{\alpha\beta}
\phi^{\beta\gamma}\phi^{\gamma\delta}\phi^{\delta\alpha}+
u_2\sum_{\alpha\beta}{\phi^{\alpha\beta}}^4\notag\\
\phantom{\frac{1}{N}\mathcal L^{(0)}}&\mathrel{\phantom{=}}
+u_3\sum_{\alpha\beta\gamma}{\phi^{\alpha\gamma}}^2
{\phi^{\beta\gamma}}^2+u_4\sum_{\alpha\beta\gamma\delta}
{\phi^{\alpha\beta}}^2 {\phi^{\gamma\delta}}^2\bigg)+\dots,\label{L0}\\
\frac{1}{\sqrt N}\mathcal L^{(1)}&=\sum_{\alpha\beta}\bigg[m\,\phi^{\alpha\beta}
-\frac{1}{2}w\sum_\gamma \phi^{\alpha\gamma}\phi^{\gamma\beta}
-\frac{1}{6}\Big(u_1\sum_{\gamma\delta}\phi^{\alpha\gamma}\phi^{\gamma\delta}
\phi^{\delta\beta}+u_2\,{\phi^{\alpha\beta}}^3+u_3\,\phi^{\alpha
\beta}\sum_\gamma {\phi^{\beta\gamma}}^2\notag\\
\phantom{\frac{1}{\sqrt N}\mathcal L^{(1)}}&\mathrel{\phantom{=}}+u_4\,\phi^{\alpha\beta}
\sum_{\gamma\delta}{\phi^{\gamma\delta}}^2\Big)+\dots\bigg]
\times\phi^{\alpha\beta}_{\mathbf p=0},\label{L1}\\
\mathcal L^{(2)}&=\frac{1}{2}\sum_{\mathbf p}\Bigg[\sum_{\alpha\beta}
\bigg(\frac{1}{2} p^2+m-\frac{1}{2}u_2{\phi^{\alpha\beta}}^2
-\frac{1}{6}u_3\sum_\gamma {\phi^{\beta\gamma}}^2
-\frac{1}{6}u_4\sum_{\gamma\delta}{\phi^{\gamma\delta}}^2+\dots
\bigg)
\times\phi^{\alpha\beta}_{\mathbf p}\phi^{\alpha\beta}_{-\mathbf p}\notag\\
\phantom{\mathcal L^{(2)}}&\mathrel{\phantom{=}}+\sum_{\alpha\beta\gamma}\bigg(
-w\phi^{\alpha\beta}-\frac{1}{3}u_1\sum_{\delta}\phi^{\alpha\delta}
\phi^{\delta\beta}-\frac{1}{3}u_3\phi^{\alpha\gamma}\phi^{\gamma\beta}
+\dots\bigg)
\times\phi^{\alpha\gamma}_{\mathbf p}\phi^{\gamma\beta}_{-\mathbf p}
\notag\\\phantom{\mathcal L^{(2)}}&\mathrel{\phantom{=}}
+\sum_{\alpha\beta\gamma\delta}\bigg(-\frac{1}{6}u_1\phi^{\alpha\gamma}
\phi^{\beta\delta}
-\frac{1}{3}u_4\phi^{\alpha\beta}\phi^{\gamma\delta}+\dots
\bigg)\times\phi^{\alpha\beta}_{\mathbf p}\phi^{\gamma\delta}_{-\mathbf p}
\Bigg],
\label{L2}\\
\sqrt N\mathcal L^{(3)}&=-\frac{1}{6}
\sideset{}{'}\sum_{\mathbf {p_1p_2p_3}}\bigg[
\sum_{\alpha\beta\gamma}\Big(w+\dots
\Big)\times
\phi^{\alpha\beta}_{\mathbf p_1}
\phi^{\beta\gamma}_{\mathbf p_2}\phi^{\gamma\alpha}_{\mathbf p_3}
+\sum_{\alpha\beta}\Big(u_2\phi^{\alpha\beta}+\dots\Big)\times
\phi^{\alpha\beta}_{\mathbf p_1}\phi^{\alpha\beta}_{\mathbf p_2}
\phi^{\alpha\beta}_{\mathbf p_3}
\notag\\
\phantom{\sqrt N\mathcal L^{(3)}}&\mathrel{\phantom{=}}
+\sum_{\alpha\beta\gamma}\Big(u_3\phi^{\beta\gamma}+\dots\Big)\times
\phi^{\alpha\beta}_{\mathbf p_1}\phi^{\alpha\beta}_{\mathbf p_2}
\phi^{\beta\gamma}_{\mathbf p_3}+
\sum_{\alpha\beta\gamma\delta}\Big(u_4\phi^{\gamma\delta}+\dots\Big)\times
\phi^{\alpha\beta}_{\mathbf p_1}\phi^{\alpha\beta}_{\mathbf p_2}
\phi^{\gamma\delta}_{\mathbf p_3}\notag\\
\phantom{\sqrt N\mathcal L^{(3)}}&\mathrel{\phantom{=}}+
\sum_{\alpha\beta\gamma\delta}\Big(u_1\phi^{\gamma\delta}+\dots\Big)\times
\phi^{\alpha\beta}_{\mathbf p_1}\phi^{\alpha\gamma}_{\mathbf p_2}
\phi^{\beta\delta}_{\mathbf p_3}+\dots\bigg].
\label{L3}
\end{align}
$\mathcal L^{(4)}$ has been omitted here, as it keeps its form in Eq.\ 
(\ref{L}) up to this order, whereas the ellipsis dots indicate the terms
higher than quartic.

\section{Perturbative calculation of the free energy}
\label{III}

Two steps lead to a systematic perturbative treatment:
\begin{itemize}
\item
Firstly an interaction Lagrangian is detached, and $\mathcal L^{(2)}$
identified as the non-interractive part:
\begin{equation}\label{Lsep}
\mathcal L=\mathcal L^{(0)}+
\frac{1}{2}\sum_{\mathbf p}\,\sum_{(\alpha\beta),(\gamma\delta)}
\big(p^2\,\delta^{\text{Kr}}_{\alpha\beta,\gamma\delta}+
\bar{M}_{\alpha\beta,\gamma\delta}\big)\,\phi^{\alpha\beta}_{\mathbf p}
\phi^{\gamma\delta}_{-{\mathbf p}}+\mathcal{L}^{\text{I}}
\end{equation}
where in the second term the bare mass is now represented by the mass
operator $\bar{M}_{\alpha\beta,\gamma\delta}$, and the summation
$\sum_{(\alpha\beta)}\equiv \sum_{\alpha<\beta}$ is over the $n(n-1)/2$
{\em pairs\/} of replica indices, i.e., over the independent field components.
Comparing with Eq.\ (\ref{L2}), the three different types of mass
components are simply derived:
\begin{align}
\bar{M}_{\alpha\beta,\alpha\beta}&=2m-\frac{1}{3}(u_1+3u_2+2u_3+
4u_4)\,{\phi^{\alpha\beta}}^2-\frac{1}{3}(2u_1+u_3+nu_4)\,
\sum_\gamma {\phi^{\beta\gamma}}^2+\dots,\notag\\
\bar{M}_{\alpha\gamma,\beta\gamma}&=-w\,\phi^{\alpha\beta}-\frac{1}{3}
(u_1+u_3+4u_4)\,\phi^{\alpha\gamma}\phi^{\beta\gamma}-\frac{1}{3}u_1\,
\sum_\delta \phi^{\alpha\delta}\phi^{\beta\delta}+\dots,\label{Mbar}\\
\bar{M}_{\alpha\beta,\gamma\delta}&=-\frac{1}{3}u_1\,
(\phi^{\alpha\gamma}\phi^{\beta\delta}+\phi^{\beta\gamma}\phi^{\alpha\delta})
-\frac{4}{3}u_4\,\phi^{\alpha\beta}\phi^{\gamma\delta}+\dots.\notag
\end{align}
Replica equivalence is a must even when replica symmetry is broken
\cite{Parisi04}, and it has been assumed in the above formulas; see the last
term of the diagonal component which is in fact independent of $\beta$.
The free propagator of this generic RSB field theory is the
$n(n-1)/2\times n(n-1)/2$ matrix $\bar G\equiv (p^2+\bar{M})^{-1}$.
\item
The order parameter $\phi^{\alpha\beta}$ must satisfy the ``equation of state"
$\langle \phi^{\alpha\beta}_{\mathbf p=0} \rangle=0$; up to one-loop order
it takes the form
\begin{equation}\label{equation_of_state}
\bar H_{\alpha\beta}+\frac{1}{2}\sum_{(\gamma\delta),(\mu\nu)}
\bar W_{\alpha\beta,\gamma\delta,\mu\nu}\times \frac{1}{N}\sum_{\mathbf p}
\bar{G}_{\gamma\delta,\mu\nu}(p)+\text{2-loop terms}=0.
\end{equation}
In perturbative computations, the capital letter representation of the
couplings is more natural, the one- and three-point ones can be deduced
by comparing with Eqs.\ (\ref{L1}) and (\ref{L3}):
\begin{equation}\label{H}
\begin{aligned}
-\bar H_{\alpha\beta}&=2\times \bigg[m\,\phi^{\alpha\beta}-
\frac{1}{2}w\sum_\gamma \phi^{\alpha\gamma}\phi^{\gamma\beta}
-\frac{1}{6}\Big(u_1\sum_{\gamma\delta}\phi^{\alpha\gamma}\phi^{\gamma\delta}
\phi^{\delta\beta}+u_2\,{\phi^{\alpha\beta}}^3+u_3\,\phi^{\alpha
\beta}\sum_\gamma {\phi^{\beta\gamma}}^2\\
\phantom{\bar H_{\alpha\beta}}&\mathrel{\phantom{=}}+u_4\,\phi^{\alpha\beta}
\sum_{\gamma\delta}{\phi^{\gamma\delta}}^2\Big)+\dots\bigg],
\end{aligned}
\end{equation}
and the eight different types of cubic couplings:
\begin{align}
\bar W_{\alpha\beta,\beta\gamma,\gamma\alpha}&=w+\dots,&
\bar W_{\alpha\beta,\alpha\beta,\alpha\beta}&=2 (u_1+u_2+u_3+2u_4)
\phi^{\alpha\beta}+\dots,\notag\\
\bar W_{\alpha\beta,\alpha\beta,\beta\gamma}&=\frac{1}{3}(2u_1+u_3+4u_4)
\phi^{\beta\gamma}+\dots,&
\bar W_{\alpha\beta,\alpha\beta,\gamma\delta}&=
\frac{4}{3}u_4\phi^{\gamma\delta}+\dots,\label{Wbar}\\[3pt]
\bar W_{\alpha\beta,\alpha\gamma,\beta\delta}&=\frac{1}{3}u_1
\phi^{\gamma\delta}+\dots\notag
\end{align}
where the dots are for the next --- $O(\phi^2)$ --- order, and the three
missing cubic vertices --- $\bar W_{\alpha\beta,\alpha\gamma,\alpha\delta}$,
$\bar W_{\alpha\gamma,\beta\gamma,\mu\nu}$, and $
W_{\alpha\beta,\gamma\delta,\mu\nu}$ --- enter only at this higher order.
\end{itemize}

The critical temperature of the field theory, or equivalently $m_c$, can be
simply deduced from Eq.\ (\ref{equation_of_state}) by 
expanding it up to its leading $O(\phi)$ order after setting $\tau=0$.
Making use of the large momentum expansion of the free propagator $\bar G$,
together with Eqs.\ (\ref{Mbar}) and (\ref{Wbar}), it straightforwardly
follows:
\begin{equation}\label{mc}
m_c=\frac{1}{2}(n-2)w^2\frac{1}{N}\sum_{\mathbf p}\frac{1}{p^4}
+\frac{1}{6}\big[(2n-1)u_1+3u_2+(n+1)u_3+(n^2-n+4)u_4\big]
\frac{1}{N}\sum_{\mathbf p}\frac{1}{p^2}+
\text{2-loop terms.}
\end{equation}
The free energy $\ln \overline{Z^n}$ has the following expansion when
$\mathcal L^{\text{I}}$ is handled as a perturbation; see (\ref{Z^n})
and (\ref{Lsep}):
\begin{equation}\label{lnZ^n}
\ln \overline{Z^n}=\ln C-\mathcal L^{(0)}+\ln Z_G-\langle L^{\text{I}}\rangle_G
+\frac{1}{2}\Big(\langle {L^{\text{I}}}^2\rangle_G-\langle L^{\text{I}}\rangle_G^2\Big)
+\dots,
\end{equation}
and the Gaussian averages $\langle\dots\rangle_G$ are taken by the measure
$e^{-\mathcal L^{(2)}}/Z_G$. As ``tadpole" diagrams are missing now due to
Eq.\ (\ref{equation_of_state}), the only one-loop term is the Gaussian
free energy $\ln Z_G= \int [d\phi]e^{-\mathcal L^{(2)}}$ which has the
familiar form in terms of the eigenvalues $\bar \lambda_j$'s of the
mass operator $\bar M$:
\begin{equation}\label{Gaussian}
\ln Z_G=-\frac{1}{2}\sum_{\mathbf p}\sum_{j=1}^{n(n-1)/2} \ln
\frac{p^2+\bar \lambda_j}{\pi}.
\end{equation}
Using the following identities:
\begin{align*}
\sum_{\mathbf p}\ln\Big(1+\frac{\bar \lambda_j}{p^2}\Big)&=
\ln\Big(1+\frac{\bar \lambda_j}{\Lambda^2}\Big)\times
\Big(\sum_{\mathbf p}1\Big)+
\bar \lambda_j\,\frac{2}{d}\,\sum_{\mathbf p}\frac{1}{p^2+\bar \lambda_j},
\\[4pt]
\frac{1}{p^2+\bar \lambda_j}&=\frac{1}{p^2}-\frac{\bar\lambda_j}{p^4}
+\frac{\bar\lambda_j^2}{p^6}
-\frac{\bar\lambda_j^3}{p^6(p^2+\bar \lambda_j)}
\end{align*}
where $\Lambda$ is the ultraviolet cutoff, and frequently applying the
trivial relationship $\sum_{\mathbf p}1=\frac{d-k}{d}\,\Lambda^k\,
\sum_{\mathbf p}\frac{1}{p^k}$ with $d>k$, the Gaussian free energy can be
arranged into its final form for $d>8$
\begin{equation}\label{lnZG}
\begin{split}
\ln Z_G &=\frac{n(n-1)}{2}\,\Big(\frac{1}{d}-\frac{1}{2}\ln \frac{\Lambda^2}{\pi}
\Big)\times \sum_{\mathbf p}1-\frac{1}{2}\Big(\sum_j \bar\lambda_j\Big)
\times \sum_{\mathbf p}\frac{1}{p^2}
+\frac{1}{4}\Big(\sum_j \bar\lambda_j^2\Big)
\times \sum_{\mathbf p}\frac{1}{p^4}\\[4pt]
&-\frac{1}{6}\Big(\sum_j \bar\lambda_j^3\Big)
\times \sum_{\mathbf p}\frac{1}{p^6}
+\frac{1}{8}\Big(\sum_j \bar\lambda_j^4\Big)\,\,\frac{d-8}{d}
\times \sum_{\mathbf p}\frac{1}{p^8}+\frac{1}{d}\sum_j\sum_{\mathbf p}
\frac{\bar\lambda_j^4}{p^6(p^2+\bar\lambda_j)}+
O(\bar\lambda_j^5)\times\! \sum_{\mathbf p}1.
\end{split}
\end{equation}

Before displaying the result for the free energy, it must be noticed that
the contribution $Nn\frac{1}{2}m_c(\phi^2)^{\alpha\alpha}$
from $\mathcal L^{(0)}$%
\footnote{Matrix notations, like $(\phi^2)^{\alpha\alpha}$ here,
are frequently used in the following part of the paper.}
is exactly cancelled by corresponding terms
of $\ln Z_G$; see Eqs.\ (\ref{L0}), (\ref{mc}) and the second terms of
the right hand sides of (\ref{TrM1}), (\ref{TrM2}).
Moreover, as $\mathcal L^{(0)}$ is stationary at the zero-loop order
parameter, we do not need to compute the one-loop correction to
$\phi^{\alpha\beta}$. Substituting the traces from App.\ \ref{App1} into
Eq.\ (\ref{lnZG}), the replicated free energy in (\ref{lnZ^n}) takes
the following form:
\begin{multline}\label{main}
\frac{1}{nN}\,\ln \overline{Z^n}=
\frac{1}{nN}\,
\big[\ln \overline{Z^n}\big]^{\text{para}}\\[5pt]
+\frac{1}{2}\tau \times (\phi^2)^{\alpha\alpha}+
\frac{1}{6}w\Big\{1-\frac{1}{2}\big[u_1+3u_2+(n-1)u_3+(n^2-n-4)u_4\big]
I_4-2(n-2)w^2I_6\Big\}
\times (\phi^3)^{\alpha\alpha}\\[5pt]
+\frac{1}{24}\Big\{u_1-\frac{1}{3}u_1\big[(2n-3)u_1+
6u_2+2(n-1)u_3+2(n^2-n-8)u_4\big]I_4-
2w^2\big[(2n-7)u_1-3u_2-(n-1)u_3\\[5pt]-(n^2-n+4)u_4\big]I_6
+3(3n-4)w^4I_8
\Big\}\times (\phi^4)^{\alpha\alpha}
+\frac{1}{24}\Big\{u_2+\frac{1}{3}
\big[2u_1^2-4(n-2)u_1u_2+3u_2^2
-2(n-5)u_2u_3\\[5pt]-2(n^2-n-8)u_2u_4+2u_3^2\big]I_4-
4w^2\big[-2u_1+(n+1)u_2+2u_3\big]I_6+24w^4I_8
\Big\}\times \sum_{\beta}{\phi^{\alpha\beta}}^4\\[1pt]
+\frac{1}{24}\Big\{(u_3+nu_4)+\frac{1}{3}
\big[(3n-4)u_1^2+12u_1u_2+8u_1u_3-(n-3)u_3^2
-2(n^2-n-8)u_3u_4
-n(n^2-n-8)u_4^2\big]I_4\\[5pt]
-2w^2\big[-(6n-19)u_1-3u_2+(n-5)u_3+(n^2-n+4)u_4\big]I_6+
3(n-12)w^4I_8
\Big\}\times \left({(\phi^2)}^{\alpha\alpha}\right)^2.
\end{multline}
Some remarks are appropriate here:
\begin{itemize}
\item
Only replica equivalence was used to derive the above formula which gives
the first order --- i.e., one-loop --- correction to the mean field free
energy. Any special RSB scheme is included in the ``invariants'' like
$(\phi^3)^{\alpha\alpha}$.
\item
For $\phi\equiv 0$, the analytic continuation of the paramagnetic free
energy is obtained:
\[
\big[\ln \overline{Z^n}\big]^{\text{para}}=\ln C-\frac{n(n-1)}{4}
\,\sum_{\mathbf p}^{\Lambda}\,\ln \frac{p^2-2\tau}{\pi};
\]
an expansion of this formula gives the $\tau$ terms in $\ln Z_G$,
see (\ref{lnZG}) and App.\ \ref{App1}. $n=1$ is the annealed model; the
disappearance of the loop corrections expresses the triviality of this case:
$\ln \overline{Z}=N\ln 2+\frac{N_B}{z}\,\frac{J^2}{2(kT)^2}$, where the
number of interactions $N_B$ can be expressed by the coordination number $z$
as $N_B=N\frac{z}{2}$. This gives for $\ln C$:
\[
\frac{1}{nN}\,\ln C=\ln 2+\frac{J^2}{4(kT)^2}
\]
which agrees with the correct value of the SK model for $n$ generic. 
\item
The notation $I_k=\frac{1}{N}\sum_{\mathbf p}^{\Lambda}\,\frac{1}{p^k}$ was used
in (\ref{main}) which is a truncated formula in the following sense:
\begin{enumerate}
\item The nonanalytic term proportional to $\tau^{d/2}$, see the last but one
loop integral in (\ref{lnZG}), was neglected, as it is subleading in the region
of space dimensions studied here, i.e.\ $d>8$.
\item Higher than quartic $\phi$ terms in (\ref{main}) were left out for the
same reason.
\item We restrict ourself to the model where quartic couplings are the highest
order, i.e.\ neglecting the invariants in (\ref{L}) which are there represented
by the dots. A fifth order coupling, for instance, would provide a contribution
$\sim I_2\times (\phi^3)^{\alpha\alpha}$ in Eq.\ (\ref{main}).
\end{enumerate}
\end{itemize}

\section{$n$-dependent free energy in special cases}
\label{IV}

Eq.\ (\ref{main}) is now used to compute $\beta\Delta\Phi\equiv
-\Big\{\frac{1}{nN}\,\ln \overline{Z^n}-
\frac{1}{nN}\,
\big[\ln \overline{Z^n}\big]^{\text{para}}\Big\}$, i.e.\ the shift of
the $n$-dependent free energy from the continuation of the paramagnetic
one, in some special cases.

\subsection{Neglecting loops: the tree approximation}

Inserting the expansions from (\ref{invariantsRSB}) into (\ref{main}),
and neglecting loop terms, the mean field free energy of the model with
nonzero $\tau$, $w$,
$u_1$, $u_2$, $u_3$, $u_4$, and assuming infinite step RSB takes the form:
\begin{equation}\label{mf_Phi}
\beta\Delta\Phi^{\text{mf}}=\frac{1}{6}w^{-2}\tau^3+\Big(\frac{1}{8}u_1
+\frac{1}{24}u_2-\frac{1}{24}u_3\Big)w^{-4}\tau^4+O(\tau^5)-\frac{1}{24}u_4
w^{-4}n\tau^4-\frac{9}{640}u_2^{-3}w^4n^5.
\end{equation}
Inserting $w=1$, $u_1=3$, $u_2=2$, $u_3=-6$, and $u_4=0$, this gives the free
energy of the SK model
\[
\beta\Phi^{\text{SK}}=-\Big[\ln 2+\frac{J^2}{4(kT)^2}\Big]+\frac{1}{6}\tau^3+
\frac{17}{24}\tau^4+O(\tau^5)-\frac{9}{5120}n^5, \qquad
2\tau=1-\left(\frac{kT}{J}\right)^2.
\]
This formula agrees with the results of Ref.\ \cite{Parisi_Rizzo_2} up to
the order studied here.\footnote{Unfortunately the definition of $\tau$
differs from that of Refs.\ \cite{Kondor,Parisi_Rizzo_1,Parisi_Rizzo_2},
which is called here $\tau'$.  The simple relation between them is
$\tau=\tau'-\frac{1}{2}{\tau'}^2$.}
From Eq.\ (\ref{mf_Phi}) follows that infinite step RSB is only a
{\em necessary\/} condition for the anomalous (non-Gaussian) free energy
fluctuation: a nonzero $u_4$ produces a term linear in $n$. This may be a
generic feature: interactions which are disconnected in replica space
(like $
\sum_{\alpha\beta\gamma\delta}
{\phi^{\alpha\beta}}^2
{\phi^{\gamma\delta}}^2$) 
generate Gaussian free energy fluctuations even at the level of the tree
approximation.

\subsection{One-loop correction for the infinite step RSB case}

When substituting $u_3$ by ${\bar u}_3\equiv u_3+nu_4$, the $n$-dependent
free energy shift takes the following simple form in the first order
perturbative calculation [see Eqs.\ (\ref{main}), (\ref{invariantsRSB}), 
(\ref{mf_Phi})]:
\[
\beta\Delta\Phi=\beta\Delta\Phi^{(0)}+n\,\beta\Delta\Phi^{(1)}
+\beta\Delta\Phi^{\text{anom}},
\]
with
\begin{align}
\beta\Delta\Phi^{(0)}&=
\frac{1}{6}w^{-2}\tau^3\big[1+(u_1+3u_2-{\bar u}_3-4u_4)I_4
-8w^2I_6\big]
+\frac{1}{24}w^{-4}\tau^4\Big[(3u_1+u_2-{\bar u}_3)\notag\\[3pt]
\phantom{\beta\Delta\Phi^{(0)}}&
\mathrel{+}\frac{1}{3}(33u_1^2+38u_1u_2-26u_1{\bar u}_3-24u_1u_4+21u_2^2-14u_2{\bar u}_3
-8u_2u_4+5{\bar u}_3^2+8{\bar u}_3u_4)I_4\notag\\[3pt]
\label{phi0}
\phantom{\beta\Delta\Phi^{(0)}}&
\mathrel{+}8(5u_1-u_2-{\bar u}_3+4u_4)\,w^2I_6+24w^4I_8\Big]
,\\[6pt]
\beta\Delta\Phi^{(1)}&=\frac{1}{6}w^{-2}\tau^3\big[{\bar u}_3I_4+4w^2I_6\big]
+\frac{1}{24}w^{-4}\tau^4\Big[\frac{1}{3}(-9u_1^2-4u_1u_2+12u_1{\bar u}_3
+8u_1u_4+4u_2{\bar u}_3\notag\\[3pt]
\phantom{\beta\Delta\Phi^{(1)}}&\mathrel{-}8u_2u_4-5{\bar u}_3^2+8u_4^2)I_4
+4(u_2+4u_4)\,w^2I_6+24w^4I_8\Big],
\label{phi1}\\[6pt]
\intertext{and}
\beta\Delta\Phi^{\text{anom}}&=-\frac{9}{640}u_2^{-4}w^4\Big\{u_2+\big[-2u_1^2
+2(2n-5)u_1u_2-9u_2^2-8u_2u_3-8u_2u_4-2u_3^2\big]I_4\notag\\[3pt]
\phantom{\beta\Delta\Phi^{\text{anom}}}&\mathrel{+}
4\big[-6u_1+(n+7)u_2+6{\bar u}_3-6nu_4\big]w^2I_6-72\,w^4I_8\Big\}
\times n^5.\label{phi_anom}
\end{align}

\subsection{For comparison: the replica symmetric free energy}
\label{RS}

The RS mean field order parameter $q\equiv \phi^{\alpha\beta}$, for any
$\alpha\not=\beta$, satisfies Eq.\ (\ref{phi}) which now takes the form
\[
\tau=-\frac{1}{2}(n-2)\,wq-\frac{1}{6}\big[(n^2-3n+3)u_1+u_2+(n-1)\bar u_3\big]
\,q^2.
\]
This equation can be used to substitute $\tau$ for $q$ in the RS relations
$\left(\phi^2\right)^{\alpha\alpha}=(n-1)\,q^2$,
$\left(\phi^3\right)^{\alpha\alpha}=(n-1)(n-2)\,q^3$, 
$\left(\phi^4\right)^{\alpha\alpha}=(n-1)(n^2-3n+3)\,q^4$,
$\sum_{\beta}{\phi^{\alpha\beta}}^4=(n-1)\,q^4$. It is now straightforward
to derive from Eq.\ (\ref{main}) the free energy shift of the replica
symmetric system with respect to the paramagnet:
\[
\beta\Delta\Phi_{\text{RS}}=\beta\Delta\Phi^{(0)}_{\text{RS}}
+n\,\beta\Delta\Phi^{(1)}_{\text{RS}}
+n^2\,\beta\Delta\Phi^{(2)}_{\text{RS}}+\dots .
\]
We can make the following observations:
\begin{itemize}
\item The RS free energy is a regular power series in $n$, with no anomalous
part, and the terms $\beta\Delta\Phi^{(k)}_{\text{RS}}$ all have the
same character as a power series of $\tau$ starting with $\tau^3$.
\item $\beta\Delta\Phi^{(0)}_{\text{RS}}=\beta\Delta\Phi^{(0)}$ of
Eq.\ (\ref{phi0}) up to $\tau^4$ (inclusively).
\item The leading term proportional to $\tau^3$ of
$\beta\Delta\Phi^{(1)}_{\text{RS}}$ is identical with the corresponding RSB
contribution in Eq.\ (\ref{phi1}). We have
\begin{align*}
\beta\Delta\Phi^{(1)}_{\text{RS}}&=
\frac{1}{6}w^{-2}\tau^3\big[{\bar u}_3I_4+4w^2I_6\big]
+\frac{1}{24}w^{-4}\tau^4\Big[u_2+\frac{1}{3}(
-7u_1^2+10u_1u_2+12u_1\bar u_3+8u_1u_4\\[3pt]
\phantom{\beta\Delta\Phi^{(1)}_{\text{RS}}}&\mathrel{+}21u_2^2+8u_2\bar u_3
-16u_2u_4-3{\bar u_3}^2+8u_4^2)\,I_4
+2(4u_1-8u_2-4\bar u_3+8u_4)\,w^2I_6+48\,w^4I_8\Big].
\end{align*}
\item From the previous observations follows that the RS and RSB free
energies differ only at the $O(5)$ order in the double series of $\tau$
and $n$.
\end{itemize}
The $O(n^2)$ term --- which is missing in the RSB scheme at one-loop
level --- is given by
\begin{equation}\label{n^2tau}
\beta\Delta\Phi^{(2)}_{\text{RS}}=
-\frac{1}{24}w^{-2}\tau^3\big[1+(u_1+3u_2-\bar u_3-4u_4)\,I_4
-8\,w^2I_6\big]+O\left(w^{-4}\tau^4\right).
\end{equation}

\section{The nonanalytic temperature dependence of the free energy}
\label{V} 

As we have seen in the previous section, the RSB free energy starts to differ
from the RS one only in the fifth analytic order $O(\tau^5)$, $O(\tau^4\,n)$,
$O(\tau^3\,n^2)$, and --- lastly but most importantly --- in the anomalous
term $O(n^5)$. In dimensions not much higher than 8, however, these terms are
subdominate with respect to the nonanalytic one proportional to $\tau^{d/2}$.
For retrieving this nonanalytic contribution from the Gaussian free energy
in Eq.\ (\ref{Gaussian}), it is converted into the following equivalent
form [Eq.\ (\ref{lnZG}) is not very useful for that purpose]:
\[
\ln Z_G=-\frac{1}{2}\sum_j\ln \frac{\Lambda^2+\bar\lambda_j}{\pi}\,
\sum_{\mathbf p}1+\frac{1}{d}\sum_j\sum_{\mathbf p}\frac{p^2}{p^2+\bar\lambda_j}
,
\]
and the last term gives the nonanalytic contribution
$\ln Z_G^{\text{na}}$ which can be generally written as
\begin{equation}\label{nonanal_lnZG}
\frac{1}{nN}\ln Z_G^{\text{na}}=
\frac{1}{d}\,
\frac{1}{N}\sum_{\mathbf p}p^2\,
\frac{1}{n}\sum_{(\alpha\beta)}\bar G_{\alpha\beta,\alpha\beta}.
\end{equation}

\subsection{$\ln Z_G^{\text{na}}$ for the infinite step RSB sheme}

At first sight, computing the right hand side of Eq.\ (\ref{nonanal_lnZG})
seems to be a formidable task due to the complicated structure of the Gaussian
propagators; see Ref.\ \cite{beyond} for details. Nevertheless, as we are
interested in the {\em leading\/} nonanalytic temperature dependence, some
important simplifications follow:
\begin{itemize}
\item
This leading term comes from the near infrared region $p^2\sim \tau$;
the propagators there
were all listed in Sec.\ 6 of Ref.\ \cite{beyond}.
When computing this near infrared propagators, the quartic couplings can be
neglected in the masses in Eq.\ (\ref{Mbar}), they enter only through the order
parameter $\phi^{\alpha\beta}$.
\item
The $n=0$ propagators --- which were computed in \cite{beyond} --- may be used, as
their $n$-dependence enters only at much higher orders. [This is just alike the
order parameter function $q(x)$.]
\item Beside the replica summations in Eq.\ \ref{nonanal_lnZG}, the origin of
the relevant $n$-dependence is $x_0$; see (\ref{n_tau}).
\end{itemize}

The trace in (\ref{nonanal_lnZG}) can be written as an integral over the continuous
overlap parameter $x$:
$\frac{1}{n}\sum_{(\alpha\beta)}\bar G_{\alpha\beta,\alpha\beta}=
-\frac{1}{2}\int_n^1 \bar G^{xx}_{11}\,dx$ and, by using Eq.\ (60)
from Ref.\ \cite{beyond}, we can write
\[
\int_n^1 p^2\,\bar G^{xx}_{11}\,dx=(1+2t+2t^2)(1-n)
-2\,\frac{1+8t+8t^2}{(1+2t)^2}\,w^2\,\frac{1}{p^4}\int_n^1q(x)^2\,dx
+\frac{8}{(1+2t)^2}\,w^4\,\frac{1}{p^8}\int_n^1q(x)^4\,dx
\]
where the dimensionless variable $t\equiv 
wq_1/p^2$  was introduced. Eq.\ (\ref{invariantsRSB}) can be used to compute
$\int_n^1q(x)^2\,dx=-\sum_{\beta}{\phi^{\alpha\beta}}^2=-\left(\phi^2\right)
^{\alpha\alpha}$, and $\int_n^1q(x)^4\,dx=-\sum_{\beta}{\phi^{\alpha\beta}}^4$.
Keeping only the relevant contributions,\footnote{Terms with an extra $x_1$
factor are subleading, and thus unimportant here. Their calculation using
the near infrared propagators from Ref.\cite{beyond} would even be
inconsistent, since corrections to these propagators give the same (subleading)
order.} we arrive at
\[
\int_n^1 p^2\,\bar G^{xx}_{11}\,dx=
1+2t-8t^3+32\,\frac{t^4(1+t)}{(1+2t)^2}-(1+2t+2t^2)\,n+\frac{81}{10}\,
\frac{1}{p^4(p^2+2wq_1)^2}\,w^8u_2^{-4}\,n^5.
\]
We are now in the position to detach the leading nonanalytic temperature 
dependence in Eq.\ (\ref{nonanal_lnZG});
approximating $wq_1\approx \tau$,
\begin{equation}\label{nonanal_RSB}
\begin{gathered}
\text{for}\quad d>8\quad \text{infinite step RSB}
:\\[10pt]
\frac{1}{nN}\ln Z_G^{\text{na}}=-\frac{16}{d}\int\limits^{\infty}
\frac{d^dp}{(2\pi)^d}\,
\frac{p^2+1}{p^6(p^2+2)^2}\,\times \tau^{d/2}
-\frac{81}{20d}\int\limits^{\infty}\frac{d^dp}{(2\pi)^d}\,
\frac{1}{p^4(p^2+2)^2}\,\times w^8u_2^{-4}\,\tau^{d/2-4}\,n^5.
\end{gathered}
\end{equation}
[Dimensional regularization is to be understood here,
allowing the ultraviolet cutoff $\Lambda$ go to infinity. This same remark is
applicable for the RS case, Eq.\ (\ref{nonanal_RS2}).]
Note the lack of the $\tau^{d/2}\,n$ term in the RSB scheme.
The $O(\tau^{d/2-4}\,n^5)$ contribution is negligible;
see Eq.\ (\ref{phi_anom}).

\subsection{$\ln Z_G^{\text{na}}$ for the RS case}

Eq.\ (\ref{nonanal_lnZG}) is now simplified as
\begin{equation}\label{nonanal_RS1}
\frac{1}{nN}\ln Z_G^{\text{na}}=
\frac{n-1}{2d}
\frac{1}{N}\sum_{\mathbf p}p^2\,
\bar G_1,
\end{equation}
with the diagonal propagator $\bar G_1$ satisfying \cite{PytteRudnick79}
\[
\frac{1}{2}n(n-1)\,\bar G_1=\frac{1}{p^2+\bar\lambda_L}+
(n-1)\,\frac{1}{p^2+\bar\lambda_A}+
\frac{1}{2}n(n-3)\,\frac{1}{p^2+\bar\lambda_R}
\]
where the three masses are in leading order
\begin{equation}\label{RSmasses}
\bar\lambda_L=2\tau,\quad\bar\lambda_A=-\frac{4}{n-2}\tau,\quad\text{and}\quad
\bar\lambda_R=-n\frac{2}{n-2}\tau.
\end{equation}
Extracting the terms providing nonanalytic temperature dependence,
Eq.\ (\ref{nonanal_RS1}) takes the following form:
\begin{equation}\label{nonanal_RS2}
\begin{gathered}
\text{for}\quad d>8\quad \text{ RS}:\\[10pt]
\frac{1}{nN}\ln Z_G^{\text{na}}=-\frac{16}{d}\int\limits^{\infty}
\frac{d^dp}{(2\pi)^d}\,
\frac{p^2+1}{p^6(p^2+2)^2}\,\times \tau^{d/2}
-\frac{8}{d}\int\limits^{\infty}\frac{d^dp}{(2\pi)^d}\,
\frac{p^2+1}{p^4(p^2+2)^3}\,\times \tau^{d/2}n.
\end{gathered}
\end{equation}
A comparison of Eqs.\ (\ref{nonanal_RSB}) and (\ref{nonanal_RS2}) shows
that the leading nonanalytic free energy contributions of the RSB and RS
phases coincide, and the free energy difference is $O(5)$ in the double
series of $\tau$ and $n$ down to dimensions 8. In the next section, the
line in the $\tau$, $n$ plane where the free energy difference disappeares
is computed in first order perturbation theory taking into account the
$O(5)$ analytical terms.

\section{The line of equal free energies of the RSB and RS schemes 
for $d>8$}
\label{VI}

To find the fifth order (in the double series of $\tau$ and $n$) results,
we must do two further steps:
\begin{itemize}
\item The one-replica quantities in Eq.\ (\ref{main}) must be extended to
the appropriate order by using the formulae in appendix \ref{App2} (for RSB)
and in subsection \ref{RS} (for RS). 
\item Eq.\ (\ref{main}) must be supplemented by the fifth order terms.
\end{itemize}

As for the first step, the notation $\delta\dots$ is introduced to mean
the difference of a one-replica quantity in the two schemes;
$\delta (\phi^2)^{\alpha\alpha}\equiv (\phi^2)^{\alpha\alpha}_{\text{RSB}}-
(\phi^2)^{\alpha\alpha}_{\text{RS}}$ for instance. Then we have
\begin{equation}\label{deltas}
\begin{aligned}
\delta (\phi^2)^{\alpha\alpha}&=
-\frac{1}{9}w^{-6}u_2^2\,\tau^4+\frac{1}{3}w^{-4}u_2\,n\tau^3
-\frac{1}{4}w^{-2}\,n^2\tau^2+O(5),\\
\delta (\phi^3)^{\alpha\alpha}&=
\frac{2}{5}w^{-7}u_2^2\,\tau^5-w^{-5}u_2\,n\tau^4+\frac{1}{2}w^{-3}\,n^2\tau^3
+\frac{27}{80}w^{3}u_2^{-3}\,n^5+O(6),\\
\delta \sum_{\beta}{\phi^{\alpha\beta}}^4&=
-\frac{8}{15}w^{-6}u_2\,\tau^5+w^{-4}\,n\tau^4
-\frac{81}{80}w^4u_2^{-4}\,n^5+O(6);
\end{aligned}
\end{equation}
whereas $\delta(\phi^4)^{\alpha\alpha}$ and
$\left({(\phi^2)}^{\alpha\alpha}\right)^2$ are of order $O(6)$.

Inclusion of the three fifth order one-replica quantities
$(\phi^5)^{\alpha\alpha}$, $\sum_{\beta}{\phi^{\alpha\beta}}^3\,
(\phi^2)^{\alpha\beta}$, and $(\phi^2)^{\alpha\alpha}\,(\phi^3)^{\alpha\alpha}$
into Eq.\ (\ref{main}) is not an easy task. Fortunately, however, the
$\delta$'s formed of them are at most of order $O(6)$, and thus negligible
for the present purpose.

We can now search for the line in the $\tau$-$n$ plane where the RS and 
infinite step RSB free energies coincide. Inserting $n=c\,w^{-2}u_2\tau$
into Eq.\ (\ref{deltas}), the following --- somewhat surprising --- result
turns up:
\begin{align*}
\delta (\phi^2)^{\alpha\alpha}&=-\frac{1}{4}(c-2/3)^2\,w^{-6}u_2^2\,\tau^4,\\
\delta (\phi^3)^{\alpha\alpha}&=\frac{9}{80}(c-2/3)^2
\,(3c^3+4c^2+4c+8)\,w^{-7}u_2^2\,\tau^5,\\
\delta \sum_{\beta}{\phi^{\alpha\beta}}^4&=-\frac{3}{80}(c-2/3)^2
\,(27c^3+36c^2+36c+32)\,w^{-6}u_2\,\tau^5.
\end{align*}
We can write the free energy difference between the RSB and RS phases up to
one-loop order, using
Eqs.\ (\ref{Phi}) and (\ref{main}), in terms of $c$ as follows:
\begin{equation}\label{Phi_vs_c}
\beta(\Phi^{\text{RSB}}-\Phi^{\text{RS}})\cong -\frac{1}{16}
\left(c-\frac{2}{3}\right)^2\,
\left[c-\frac{2}{3}-2u_2^{-1}f_d(\Lambda)\right]\,w^{-6}u_2^2\,\tau^5,
\qquad c\to \frac{2}{3}
\end{equation}
where the correction term is
\begin{equation}\label{f}
f_d(\Lambda)=\frac{1}{3}(2u_1^2+11u_1u_2+12u_2^2+7u_2u_3
+4u_2u_4+2u_3^2)\,I_4+4(2u_1-3u_2-2u_3)w^2I_6+24w^4I_8.
\end{equation}
It can be seen from Eq.\ (\ref{Phi_vs_c}) that
for the cases $c\approx 2/3$, the RSB and RS free energies
differ only at 3-loop order! Without any further assumption, we should go
up to that order to gain the 1-loop correction to $c_{\text{eq}}$,
i.e.\ to the $c$ value of the line where
the two free energies are equal. We will, however, assume that for $d>8$ the
mean field form of Eq.\ (\ref{Phi_vs_c}), i.e.
\begin{equation}\label{condition}
\beta(\Phi^{\text{RSB}}-\Phi^{\text{RS}})\sim (c-c_{\text{eq}})^3,
\end {equation}
remains valid for the model. Eq.\ (\ref{Phi_vs_c}) then provides
$c_{\text{eq}}=\frac{2}{3}[1+u_2^{-1}f_d(\Lambda)]$, and finally
\begin{equation}\label{result}
\text{for $d>8$}\quad\Phi^{\text{RSB}}=\Phi^{\text{RS}}\Longrightarrow
n=\frac{2}{3}\left[u_2+f_d(\Lambda)+O(\text{2-loop})\right]\,w^{-2}\tau.
\end{equation}

\section{Between six and eight dimensions}
\label{VII}

\subsection{The infinite step RSB case}
As it has been explained in Ref.\ \cite{beyond}, the quartic vertex
whose bare value is $u_2$%
\footnote{In \cite{beyond} $u=u_2/2$ was used.}
suffers a change in its temperature scaling
from $\tau^0$ to $\tau^{d/2-4}$ when crossing $d=8$. The RSB order parameter
is severely influenced by this, as best seen from Eqs.\
(\ref{equation_of_state}), (\ref{H}) by extracting a term proportional to $q(x)^3$
from the one-loop contribution, see \cite{beyond}, and matching it to the
corresponding zero-loop one:
\[
\frac{1}{3}\Big[u_2+24
\int\limits^{\infty}\frac{d^dp}{(2\pi)^d}\,
\frac{1}{p^4(p^2+2)^2}\,w^4 \tau^{d/2-4}\Big]\times
{\phi^{\alpha\beta}}^3.
\]
The second part between the brackets is called $\tilde u_2$ and reproduced
in Appendix \ref{App3}; it obviously dominates the bare value $u_2$.
Beside $\tilde u_2$, the combination $\tilde u_1-\frac{1}{3}\tilde {\bar u}_3$
emerges too from the one-loop correction, as shown in Appendix \ref{App3},
and the equation of state in (\ref{equation_of_state}) and (\ref{H}),
with the relevant terms kept only, takes the following form [the usual
infinite step ultrametric ansatz is assumed here, and therefore $q(x)$
replaces $\phi^{\alpha\beta}$]:
\begin{equation}\label{6<d<8_equation_of_state}
2\tau q(x)+w\left[\Big(n-\frac{2}{3}x_0\Big)q_0^2
-2\Big(q_1-\frac{1}{2}x_1q_1\Big)q(x)
-\frac{1}{3}xq(x)^2 \right]
+\left(\tilde u_1-\frac{1}{3}\tilde {\bar u}_3\right) q_1^2q(x)
+\frac{1}{3}\tilde u_2 q(x)^3=0.
\end{equation}
Solving this equation provides the order parameter function in this
transitional domain of dimensions:
\begin{equation}\label{tilded_order_parameter}
q(x)=\frac{w}{\tilde u_2}\,x,\qquad\tau=wq_1-\frac{1}{2}\left(\tilde u_1
+\tilde u_2-\frac{1}{3}\tilde {\bar u}_3\right)q_1^2,\qquad
n=\frac{2}{3}x_0,\qquad 6<d<8;
\end{equation}
compare it with (\ref{q(x)}), (\ref{K}), and (\ref{n_tau}). The tilded
quantities are proportional to $w^4\,\tau^{d/2-4}$, see Eq.\
(\ref{tilded_quantities}).
Note that although $x_1\cong\frac{\tilde u_2}{w^2}\,\tau\sim w^2\tau^{d/2-3}$
is one-loop order, the order parameter $q(x)$ continues to be zero-loop
order.

We are now proceeding to compute the relevant nonanalytic terms to
the free energy in (\ref{lnZ^n}): both $\mathcal L^{(0)}$ and
$\ln Z_G$ must be considered in this regime of dimensions.
\begin{itemize}
\item
Gaussian free energy: near infrared contribution.
Eq.\ (\ref{nonanal_RSB}) remains valid, except replacing $u_2$ by $\tilde
u_2$ in the second term, and exploiting the definition of $\tilde u_2$
in (\ref{tilded_quantities}):
\begin{equation}\label{tilded_n^5_1}
\frac{1}{nN}\ln Z_G^{\text{na}}=-\frac{16}{d}\int\limits^{\infty}
\frac{d^dp}{(2\pi)^d}\,
\frac{p^2+1}{p^6(p^2+2)^2}\,\times \tau^{d/2}
-\frac{27}{160d}\,w^4\tilde u_2^{-3}n^5.
\end{equation}
\item
A detailed analysis based on Ref.\ \cite{beyond} shows
that only highly subdominant contributions
--- $\frac{1}{nN}\ln Z_G^{\text{na}}\sim\tau^{d-1}$ --- result from
the far infrared regime $p^2\sim u_2q^2\sim \frac{u_2}{w^2}\tau^2$; these are
even smaller than the correction term from the near infrared regime which is
$\sim \tau^{d/2+1}$.
\item
Nonanalytic temperature dependence occurs in $\mathcal L^{(0)}$ of Eq.\
(\ref{L0}) due to the tilded quartic couplings in the order parameter,
see (\ref{tilded_order_parameter}). For retrieving the relevant leading
contributions, it is sufficient to take
\[
-\frac{1}{nN}\mathcal L^{(0)}=\frac{1}{2}\tau (\phi^2)^{\alpha\alpha}
+\frac{1}{6}w(\phi^3)^{\alpha\alpha},
\]
thus neglecting terms which are smaller by $u_2/\tilde u_2\sim \tau^{4-d/2}$.
A straightforward calculation provides
\begin{equation}\label{tilded_n^5_2}
-\frac{1}{nN}\mathcal L^{(0)}=C_{\text{2-loop}}\times w^2\tau^{d-3}+
\frac{9}{160}\,w^4\tilde u_2^{-3}n^5.
\end{equation}
(Note the lack of the term $\sim\tau^{d/2}$.) The first term, which
obviously dominates the subleading one $\sim\tau^{d/2+1}$, consists of
contributions like $w^{-1}\tilde u_2^2q_1^5$. A consistent calculation of
$C_{\text{2-loop}}$, however, requires the 2-loop extension of the equation
of state. Similarly, 2-loop corrections behind the Gaussian free energy
in Eq.\ (\ref{lnZ^n}) yield the same kind of term.
\end{itemize}

\subsection{The replica symmetric case}

Similarly to the RSB case, we must include the one-loop term in
Eq.\ (\ref{equation_of_state}), and --- keeping only the relevant terms
--- the RS equation of state reads:
\[
2\tau q+(n-2)wq^2+(n-2)w\frac{1}{N}\sum_{\mathbf p}\bar G_2=0,
\]
with \cite{PytteRudnick79}
\[
n(n-1)(n-2)\bar G_2=2(n-2)\,\frac{1}{p^2+\bar\lambda_L}+
(n-1)(n-4)\,\frac{1}{p^2+\bar\lambda_A}-n(n-3)\,\frac{1}{p^2+\bar\lambda_R},
\]
and for the RS masses see (\ref{RSmasses}).
Extracting the nonanalytic contribution from the third term is somewhat
lengthy but straightforward; in the zero replica number limit it can
be put into the form [see Eq.\ (\ref{tilded_quantities})]
\[
(n-2)w\frac{1}{N}\sum_{\mathbf p}\bar G_2=\left[\left(\tilde u_1
-\frac{1}{3}\tilde {\bar u}_3\right)+\frac{1}{3}\tilde u_2
\right]\,q^3,\qquad n=0.
\]
Note that the resultant RS equation of state is equivalent with the
RSB one of Eq.\ (\ref{6<d<8_equation_of_state}) for $x=x_0=x_1$.

We are now in the position to analyse the two possible sources for the RS
free energy having nonanalytic temperature dependence for $6<d<8$:
\begin{itemize}
\item As for the Gaussian part, Eq.\ (\ref{nonanal_RS2}) remains valid
for $\frac{1}{nN}\ln Z_G^{\text{na}}$ even in this dimensional regime.
\item Using the RS equation of state from above, it is straightforward
to see that the leading nonanalytic term to $-\frac{1}{nN}\mathcal L^{(0)}$
is proportional to $\tau^{d-3}$. It has, however, a 2-loop character, and
a consistent calculation would require extending to the next perturbative
order.
\end{itemize}

\subsection{The free energy difference between the RSB ans RS phases}

Collecting pieces of information from previous sections and subsections,
namely Eqs.\ (\ref{n^2tau}), (\ref{nonanal_RS2}), (\ref{tilded_n^5_1})
and (\ref{tilded_n^5_2}), the leading terms which do not cancel
in the free energy difference when $6<d<8$ are the following:
\begin{multline*}
\beta(\Phi^{\text{RSB}}-\Phi^{\text{RS}})=\\
C_{\text{2-loop}}\times w^2\tau^{d-3}
+\frac{8}{d}\int\limits^{\infty}\frac{d^dp}{(2\pi)^d}\,
\frac{p^2+1}{p^4(p^2+2)^3}\,\times \tau^{d/2}n
+\frac{1}{24}w^{-2}\tau^3n^2
+\frac{9}{160d}(3-d)\,w^4\tilde u_2^{-3}n^5.
\end{multline*}
[The notation $C_{\text{2-loop}}$ was kept here for simplicity, although
it may obviously differ from that defined in Eq.\ (\ref{tilded_n^5_2}).
For the definition of $\tilde u_2$ see (\ref{tilded_quantities}).]
The line where the RSB and RS free energies coincide follows from this
equation:
\[
n\sim w^2\,\tau^{d/2-3},\qquad 6<d<8.
\]
The proportionality constant is a one-loop integral, whose value can only
be computed from the knowledge of the two-loop integral $C_{\text{2-loop}}$.

\section{The stability boundary of the RS phase}
\label{VIII}

It is well known from the famous finding of de Almeida and Thouless
\cite{AT} that the mean field Ising spin glass in a homogeneous magnetic
field enters the RSB phase along the boundary where the RS phase becomes
unstable. This was later extended by Kondor \cite{Kondor} to the case
when the magnetic field is zero, the replica number $n$, however, is 
finite: $n$ essentially takes over the role of the magnetic field, and
along an Almeida-Thouless (AT) line in the $\tau-n$ plane, the RS phase becomes
unstable. In \cite{Kondor} a simplified model --- the ``truncated" model with
all the quartic couplings but $u_2$ zero --- was used, and it was shown
that the RS and RSB free energies coincide along the instability line.

In Ref.\ \cite{droplet} the leading behaviour of the dangerous ``replicon"
mass $\Gamma_\text{R}$ close to $T_c$ was expressed in terms of 
the {\em exact\/} cubic and
quartic vertices $w^{\text{exact}}$ and $u_2^{\text{exact}}$, and also the
exact order parameter $q^{\text{exact}}$ as%
\footnote{The superscript ``exact" is used to distinguish these quantities
from the ``bare" ones, which are the zeroth order contributions to them.}
\[
\Gamma_\text{R}=nw^{\text{exact}}q^{\text{exact}}-\frac{2}{3}u_2^{\text{exact}}
{q^{\text{exact}}}^2,\qquad d>8.
\]
The zero replicon mode signals the instability of the RS phase providing
for the AT-line:
\[
n_{\text{AT}}=\frac{2}{3}{w^{\text{exact}}}^{-1}u_2^{\text{exact}}
q^{\text{exact}}.
\]
Substituting the bare values for the vertices and the mean field order
parameter $wq=\tau$, agreement with the zero-loop result for the line of
equal free energies of Eq.\ (\ref{result}) is found, thus reproducing
Kondor's result. (Although the more generic model with all the quartic
couplings is considered here.)

Computing the one-loop correction of the replicon mass by standard
perturbative methods is straightforward for $d>8$, although somewhat lengthy.
Omitting the details, only the final result is displayed here:
\begin{multline*}
\Gamma_\text{R}=\left\{1+[(n-1)u_1+u_3+4u_4]\,I_4+(n-2)w^2\,I_6\right\}\,n\,
(wq^{\text{exact}})+
\bigg\{\frac{1}{3}[n(n-3)u_1-2u_2]\\[3pt]
+\frac{1}{9}[(2n^3-5n^2-3n-4)u_1^2
-12u_1u_2+4n(n-3)u_1u_3+24n(n-3)u_1u_4-18u_2^2-24u_2u_3-48u_2u_4-4u_3^2]\,I_4\\[5pt]
+\frac{1}{3}[-(2n^3-6n^2+12n+16)u_1+24u_2-2(n^2-8)u_3+8n(n-6)u_4]\,w^2I_6
-(2n^3-9n^2+12n+16)\,w^4I_8
\bigg\}\,{q^{\text{exact}}}^2.
\end{multline*}
This must be complemented by the equation of state, i.e.\ by the one-loop
relationship between the order parameter and $\tau$:
\begin{multline*}
-2\tau=(n-2)\left\{1
+\frac{1}{3}[(n-2)u_1-3u_2-(n-2)u_3-(n^2-n-8)
u_4]\,I_4-(n-2)w^2I_6\right\}(wq^{\text{exact}})\\[3pt]
+O({q^{\text{exact}}}^2).
\end{multline*}
From these two equations, we can easily conclude the leading behaviour for
$n\sim \tau \ll 1$:
\begin{multline}\label{Rmass}
\Gamma_\text{R}=\left[1+\frac{1}{3}(-u_1+3u_2+u_3+4u_4)\,I_4-4w^2I_6\right]\,n\tau
-\frac{2}{3}\bigg[u_2+\frac{1}{3}(2u_1^2+10u_1u_2+15u_2^2\\[4pt]
+8u_2u_3+8u_2u_4
+2u_3^2)\,I_4+8(u_1-2u_2-u_3)w^2I_6+24w^4I_8\bigg]\,w^{-2}\tau^2.
\end{multline}
The instability line is obtained from the condition $\Gamma_\text{R}=0$:
\[
n_{\text{AT}}=\frac{2}{3}[u_2+f_d(\Lambda)]\,w^{-2}\tau
\]
where the correction term has been defined in (\ref{f}).
Comparing this expression with (\ref{result}), we can conclude that the mean
field type behaviour persists for $d>8$: the RS phase becomes unstable where
its free energy coincides with that of the RSB phase. (See, however, the
discussion in the Conclusion part.)

\section{Conclusion}
\label{IX}

Two basic features of the mean field Ising spin glass were followed in this
paper, by perturbatively taking into account the effect of the geometry
of a high dimensional lattice ($d$ certainly larger than 6).
These properties are:
\begin{itemize}
\item the anomalous sample to sample free energy fluctuations (considering
only {\em large\/} deviations),
\item and the equality of the free energies of the replica symmetric and
infinite step replica symmetry broken phases along the line (Almeida-Thouless
line) where the RS phase becomes unstable.
\end{itemize}
Both problems can be elaborated by studying the $n$-dependent free energy
below the spin glass transition.

As for the first item, perturbations break down the anomalous behaviour, and
Gaussian large deviations take over the lead. As it was shown in Sec.\
\ref{IV}, Gaussian fluctuation is common for {\em any\/} ansatz of the
order parameter with the property of replica eqivalence (not to be confused
with replica symmetry), i.e.\ it must be a geometrical effect.
As it was pointed out by G.\ Parisi \cite{private_communication_Parisi1},
Gaussian fluctuations always dominate whenever local interactions are
inhomogeneous: this is certainly the case in the finite dimensional geometry
of a hypercubic lattice (but not for the SK model).
The fact that locally inhomogeneous interactions
imply Gaussian large deviations
of the free energy has been demonstrated recently on the Bethe lattice with
finite connectivity, and continuously distributed quenched interactions
\cite{Parisi_Rizzo_3}.
The anomalous
$n$-dependence of the RSB free energy, however, does persist, although it
is subleading in finite dimensions.

The coincidence of the RS instability line and the line of equal free energies
with the RSB phase is somewhat misterious even in the SK model: the free
energy difference is of fifth order (in the double series in $\tau$ and
$n$), as contrasted with the replicon mass wich is a second order quantity.
We found that this feature of the mean field theory persists for $d>8$,
but in a rather nontrivial way (see the complicated correction term in
Eq.\ (\ref{f})). We must emphasize that the two computations in Secs.\
\ref{VI} and \ref{VIII} are completely independent. This result gives an
important support to the scenario, at least for $d>8$,
that replica symmetry must be broken by the infinite
ultrametric hierarchy of the mean field spin glass proposed by Parisi.
We must remember, however, that the condition of Eq.\ (\ref{condition})
was a priori assumed, and any conclusions depend on the validity of it.

In the dimensional domain $6<d<8$ the situation is more complicated.
The AT line can be computed relatively easily, as it was in Ref.\
\cite{nucl}:
\[
n_{\text{AT}}=\frac{2}{3}w^{-2}\tilde u_2\tau \sim w^2\tau^{d/2-3},
\]
see (\ref{tilded_quantities}) and Eq.\ (47) of \cite{nucl}.
Among the leading contributions to the free energy difference, which are now
nonanalytic, there are terms which can be computed only by extending
the calculation to two-loop order; as explained in Sec.\ \ref{VII}.
This seems to be unfeasible, and no a priori assumptions are available now.
A real miracle would be the coincidence of the two lines, presumably
due to several cancellations. One can speculate that
otherwise the separation of the two lines
might be explained by some non mean field scenarios: first order transition
or replica symmetry breaking with replica equivalence, but not with the
infinite step ultrametric structure. 

\begin{acknowledgments}
This work has benefited from
a useful correspondence with Giorgio Parisi and Tommaso Rizzo.
Sending me preliminary results of Ref.\ \cite{Parisi_Rizzo_3} is also
acknowledged.
\end{acknowledgments}

\appendix
\section{Traces of powers of the Gaussian mass operator
for a generic RSB scheme with replica equivalence}\label{App1}

Traces appearing in the formula for $\ln Z_G$, Eq.\ (\ref{lnZG}),
can be computed and arranged into the generic form
\[
\frac{1}{n}\,\sum_j {\bar \lambda}^k_j=\frac{1}{n}\,\text{Tr}\,{\bar M}^k
\sim \quad\tau^k\quad
+\quad\Big[\text{expression of}\quad(\phi^2)_,^{\alpha\alpha}\,\,
(\phi^3)_,^{\alpha\alpha}\,\,(\phi^4)_,^{\alpha\alpha}\,\,\,
\sum_{\beta}{\phi^{\alpha\beta}}^4,\dots\Big].
\]
These formulae are derived by two subsequent steps:
\begin{itemize}
\item Substitution of the mass components from (\ref{Mbar}).
\item From the zero-loop equation of state ${\bar H}_{\alpha\beta}=0$,
with $m=-\tau$ in (\ref{H}),
and exploiting replica equivalence new,
replica {\em independent\/} equations follow, like this:
\[
\tau\,(\phi^2)^{\alpha\alpha}=-\frac{1}{2}w\,(\phi^3)^{\alpha\alpha}
-\frac{1}{6}
\Big[u_1\,{(\phi^4)}^{\alpha\alpha}+u_2\,\sum_\beta{\phi^{\alpha\beta}}^4
+(u_3+nu_4)\,\left({(\phi^2)}^{\alpha\alpha}\right)^2\Big].
\]
\end{itemize}
In this way, we can get rid of terms where $\tau$ and $\phi$ are coupled.

The following formulae are valid for any RSB scheme compatible with
replica equivalence, not necessarily ultrametric:
\begin{equation}\label{TrM1}
\frac{1}{n}\text{Tr}\,{\bar M}
=-(n-1)\tau-\frac{1}{6}\big[(2n-1)u_1+3u_2+(n+1)u_3
+(n^2-n+4)u_4\big]\,(\phi^2)^{\alpha\alpha};
\end{equation}
\begin{multline}\label{TrM2}
\frac{1}{n}\text{Tr}\,{\bar M}^2=2(n-1)\tau^2+(n-2)w^2\,(\phi^2)^{\alpha\alpha}
-\frac{1}{3}w\big[u_1+3u_2+(n-1)u_3+(n^2-n-4)u_4\big]\,
(\phi^3)^{\alpha\alpha}\\[5pt]
-\frac{1}{18}u_1\big[(2n-3)u_1+6u_2+2(n-1)u_3+2(n^2-n-8)u_4\big]\,
{(\phi^4)}^{\alpha\alpha}+
\frac{1}{18}\big[2u_1^2-4(n-2)u_1u_2+3u_2^2\\[5pt]
-2(n-5)u_2u_3
-2(n^2-n-8)u_2u_4+
2u_3^2\big]\,\sum_{\beta}{\phi^{\alpha\beta}}^4+
\frac{1}{18}\big[(3n-4)u_1^2+12u_1u_2+8u_1u_3-(n-3)u_3^2\\[1pt]
-2(n^2-n-8)u_3u_4
-n(n^2-n-8)u_4^2\big]\,\left({(\phi^2)}^{\alpha\alpha}\right)^2;
\end{multline}
\begin{multline*}
\frac{1}{n}\text{Tr}\,{\bar M}^3=-4(n-1)\tau^3+2(n-2)w^3\,(\phi^3)^{\alpha\alpha}
+\frac{1}{2}w^2\big[(2n-7)u_1-3u_2-(n-1)u_3-(n^2-n+4)u_4\big]\,
{(\phi^4)}^{\alpha\alpha}\\[5pt]
+w^2\big[-2u_1+(n+1)u_2+2u_3\big]\,
\sum_{\beta}{\phi^{\alpha\beta}}^4+
\frac{1}{2}w^2\big[-(6n-19)u_1-3u_2+(n-5)u_3+\\(n^2-n+4)u_4\big]\,
\left({(\phi^2)}^{\alpha\alpha}\right)^2+\dots;
\end{multline*}
\[
\frac{1}{n}\text{Tr}\,{\bar M}^4=8(n-1)\tau^4+w^4\Big[(3n-4)\,
{(\phi^4)}^{\alpha\alpha}+8\,\sum_{\beta}{\phi^{\alpha\beta}}^4+
2(n-7)\,\left({(\phi^2)}^{\alpha\alpha}\right)^2\Big]+\dots.
\]

\section{Zero-loop order parameter of the $\tau$, $u_1$,
$u_2$, $u_3$, $u_4$ model}\label{App2}

For the model with higher than quartic invariants neglected --- which is
somewhat more general than the traditional ``truncated" model where
$u_1=u_3=u_4=0$ ---, the mean field equation of state follows from Eqs.\
(\ref{equation_of_state}) and (\ref{H}) by neglecting the 1-loop term
and setting $m=-\tau$:
\begin{equation}\label{phi}
\tau\,\phi^{\alpha\beta}=
-\frac{1}{2}w\, {(\phi^2)}^{\alpha\beta}
-\frac{1}{6} \big[u_1\,{(\phi^3)}^{\alpha\beta}+u_2\,{\phi^{\alpha\beta}}^3
+(u_3+nu_4)\,{(\phi^2)}^{\alpha\alpha}\phi^{\alpha\beta}\big] .
\end{equation}
(Replica equivalence was
applied to get the last term.) The infinite-step ultrametric solution 
of this equation
can be represented by the order parameter {\em function\/} $q(x)$,
see \cite{MePaVi} and references therein, and $x$ falls into the interval
$[n,1]$ for $n>0$ small but finite. An application of the ultrametric
replica algebra \cite{MePa}, and extending the method used for the
truncated model \cite{Parisi80_truncated}, a somewhat lengthy calculation
leads to
\begin{equation}\label{q(x)}
q(x)=
\begin{cases}
q_0 & n\le x\le x_0\\
K\,\frac{x}{\sqrt{1+\frac{u_1}{u_2}x^2}} & x_0\le x\le x_1\\
q_1 & x_1\le x\le 1,
\end{cases}
\end{equation}
and the constant $K$ is best expressed by $x_1$ as
\begin{equation}\label{K}
K=\frac{w}{u_2}\, \frac{\sqrt{1+\frac{u_1}{u_2}x_1^2}}{1+\frac{u_1}{u_2}x_1}.
\end{equation}
Continuity of $q(x)$ at $x_0$ and $x_1$ means that only two of the four
parameters $x_0$, $x_1$, $q_0$, and $q_1$ are independent. Choosing
$x_0$ and $q_1$ as such, the most economic way to give them as function
of the model parameters is the inverse relationship:
\begin{equation}\label{n_tau}
\begin{aligned}
n&=\frac{2}{3}\,x_0+O(3),\\
\tau&=wq_1-\frac{1}{2}\Big[u_1+u_2-\frac{1}{3}(u_3+nu_4)\Big]q_1^2
-\frac{1}{9}\frac{u_2u_3}{w}q_1^3+O(4).
\end{aligned}
\end{equation}
[Orders $O(\dots)$ are understood in the sense of the double series
in $x_0$ and $q_1$.
Notice the lack of $O(2)$ term in the first equation.]

The invariants entering Eq.\ (\ref{main}) can be computed by the methods
presented in \cite{MePa}, and they have the following expansions:
\begin{equation}\label{invariantsRSB}
\begin{aligned}
&(\phi^2)^{\alpha\alpha}=-q_1^2+\frac{2}{3}x_1q_1^2+O(5)=
-w^{-2}\Big[\tau^2+\frac{1}{3}w^{-2}\big(3u_1+u_2-u_3-nu_4\big)\tau^3
+\dots\Big],\\
&(\phi^3)^{\alpha\alpha}=2q_1^3-2x_1q_1^3+\frac{2}{5}x_1^2q_1^3+\frac{2}{45}
x_0^2q_0^3+O(6)=w^{-3}\Big[2\tau^3+w^{-2}\big(3u_1+u_2-
u_3-nu_4\big)\tau^4+\dots\Big]\\
&\phantom{(\phi^3)^{\alpha\alpha}}\mathrel{\phantom{=}}+\frac{27}{80}w^3u_2^{-3}\,n^5,\\
&(\phi^4)^{\alpha\alpha}=-3q_1^4+4x_1q_1^4+O(6)=-3w^{-4}\tau^4+\dots,\\[6pt]
&\sum_{\beta}{\phi^{\alpha\beta}}^4=-q_1^4+\frac{4}{5}x_1q_1^4
-\frac{2}{15}x_0q_0^4+O(6)=-w^{-4}\tau^4+\dots-\frac{81}{80}
w^{4}u_2^{-4}\,n^5
,\\[2pt]
&\left({(\phi^2)}^{\alpha\alpha}\right)^2=q_1^4-\frac{4}{3}x_1q_1^4+O(6)=
w^{-4}\tau^4+\dots.
\end{aligned}
\end{equation}
While deriving the above expansions, Eqs.\ (\ref{q(x)}), (\ref{K})
and (\ref{n_tau}) were frequently used.

\section{Calculation of the nonanalytic contributions for the equation of
state}\label{App3}

Extracting the {\em leading\/} nonanalytic contributions from the second
term in Eq.\ (\ref{equation_of_state}) becomes possible by the following
simplifications:
\begin{itemize}
\item All the bare cubic vertices but
$\bar W_{\alpha\beta,\beta\gamma,\gamma\alpha}=w$ are negligible, see
Eq.\ (\ref{Wbar}).
\item We can use the bare propagators in the near infrared region
\cite{beyond}; this is equivalent by neglecting the bare quartic couplings
$u_1$, $u_2$, $u_3$, and $u_4$ in the mass components in (\ref{Mbar}).
\item The single replica sum arising, see below, is then approximated by
evaluating it in the smallest ultrametric block of size $x_1-1$, and $x_1$
is obviously negligible here for small $\tau$ and $d>6$.
\end{itemize}
We can write then
\[
\frac{1}{2}\sum_{(\gamma\delta),(\mu\nu)}
\bar W_{\alpha\beta,\gamma\delta,\mu\nu}
\,\bar{G}_{\gamma\delta,\mu\nu}
\cong w \sum_{\gamma\not=\alpha\beta} {\bar G}_{\alpha\gamma,\beta\gamma}
\cong -2w \,{\bar G}^{xx_1}_{1x},
\]
$x$ being the ultrametric distance between $\alpha$ and $\beta$, and the
parametrization of ultrametric matrices was used in the last formula
\cite{beyond}.
Eq.\ (62) of Ref.\ \cite{beyond} with $y=x_1$ can now be used:
\[
p^2\,{\bar G}^{xx_1}_{1x}=\frac{(1+6t+4t^2-4t_x^2)\,t_x}{(1+2t)^2}=
t_x+2tt_x-8\frac{(1+t)\,t^2t_x}{(1+2t)^2}-4\frac{t_x^3}{(1+2t)^2}
\]
where $t\equiv wq_1/p^2$ and $t_x\equiv wq(x)/p^2$.
Only the last two terms contribute to the nonanalytic temperature
dependence we are looking for: the second term in Eq.\ (\ref{equation_of_state})
can then be written as
\[
\left(\tilde u_1-\frac{1}{3}\tilde {\bar u}_3\right)
\,q_1^2q(x)+\frac{1}{3}\tilde u_2\,q(x)^3
\]
with
\begin{equation}\label{tilded_quantities}
\begin{aligned}
\tilde u_1-\frac{1}{3}\tilde {\bar u}_3&=
16\int\limits^{\infty}\frac{d^dp}{(2\pi)^d}\,\frac{p^2+1}{p^6(p^2+2)^2}\,w^4\,
\tau^{d/2-4},\\[6pt]
\tilde u_2&=24\int\limits^{\infty}\frac{d^dp}{(2\pi)^d}\,\frac{1}{p^4(p^2+2)^2}
\,w^4\,\tau^{d/2-4}.
\end{aligned}
\end{equation}

\begin{thebibliography}{20}
\expandafter\ifx\csname natexlab\endcsname\relax\def\natexlab#1{#1}\fi
\expandafter\ifx\csname bibnamefont\endcsname\relax
  \def\bibnamefont#1{#1}\fi
\expandafter\ifx\csname bibfnamefont\endcsname\relax
  \def\bibfnamefont#1{#1}\fi
\expandafter\ifx\csname citenamefont\endcsname\relax
  \def\citenamefont#1{#1}\fi
\expandafter\ifx\csname url\endcsname\relax
  \def\url#1{\texttt{#1}}\fi
\expandafter\ifx\csname urlprefix\endcsname\relax\def\urlprefix{URL }\fi
\providecommand{\bibinfo}[2]{#2}
\providecommand{\eprint}[2][]{\url{#2}}

\bibitem[{\citenamefont{Crisanti et~al.}(1992)\citenamefont{Crisanti, Paladin,
  Sommers, and Vulpiani}}]{Crisanti_et_al}
\bibinfo{author}{\bibfnamefont{A.}~\bibnamefont{Crisanti}},
  \bibinfo{author}{\bibfnamefont{G.}~\bibnamefont{Paladin}},
  \bibinfo{author}{\bibfnamefont{H.-J.} \bibnamefont{Sommers}},
  \bibnamefont{and} \bibinfo{author}{\bibfnamefont{A.}~\bibnamefont{Vulpiani}},
  \bibinfo{journal}{J. Phys. I (France)} \textbf{\bibinfo{volume}{2}},
  \bibinfo{pages}{1325} (\bibinfo{year}{1992}).

\bibitem[{\citenamefont{Parisi and Rizzo}(2008)}]{Parisi_Rizzo_1}
\bibinfo{author}{\bibfnamefont{G.}~\bibnamefont{Parisi}} \bibnamefont{and}
  \bibinfo{author}{\bibfnamefont{T.}~\bibnamefont{Rizzo}},
  \bibinfo{journal}{Phys. Rev. Lett.} \textbf{\bibinfo{volume}{101}},
  \bibinfo{pages}{117205} (\bibinfo{year}{2008}).

\bibitem[{\citenamefont{Parisi and Rizzo}(2009{\natexlab{a}})}]{Parisi_Rizzo_2}
\bibinfo{author}{\bibfnamefont{G.}~\bibnamefont{Parisi}} \bibnamefont{and}
  \bibinfo{author}{\bibfnamefont{T.}~\bibnamefont{Rizzo}},
  \bibinfo{journal}{Phys. Rev. B} \textbf{\bibinfo{volume}{79}},
  \bibinfo{pages}{134205} (\bibinfo{year}{2009}{\natexlab{a}}),
  \eprint{arXiv:0811.1524}.

\bibitem[{\citenamefont{Kondor}(1983)}]{Kondor}
\bibinfo{author}{\bibfnamefont{I.}~\bibnamefont{Kondor}}, \bibinfo{journal}{J.
  Phys. A} \textbf{\bibinfo{volume}{16}}, \bibinfo{pages}{L127}
  (\bibinfo{year}{1983}).

\bibitem[{\citenamefont{Sherrington and Kirkpatrick}(1975)}]{SK}
\bibinfo{author}{\bibfnamefont{D.}~\bibnamefont{Sherrington}} \bibnamefont{and}
  \bibinfo{author}{\bibfnamefont{S.}~\bibnamefont{Kirkpatrick}},
  \bibinfo{journal}{\prl} \textbf{\bibinfo{volume}{35}}, \bibinfo{pages}{1792}
  (\bibinfo{year}{1975}).

\bibitem[{\citenamefont{Bray and Moore}(1979)}]{BrMo79}
\bibinfo{author}{\bibfnamefont{A.~J.} \bibnamefont{Bray}} \bibnamefont{and}
  \bibinfo{author}{\bibfnamefont{M.~A.} \bibnamefont{Moore}},
  \bibinfo{journal}{J. Phys. C} \textbf{\bibinfo{volume}{12}},
  \bibinfo{pages}{79} (\bibinfo{year}{1979}).

\bibitem[{\citenamefont{Temesv{\'a}ri et~al.}(2002)\citenamefont{Temesv{\'a}ri,
  De~Dominicis, and Pimentel}}]{rscikk}
\bibinfo{author}{\bibfnamefont{T.}~\bibnamefont{Temesv{\'a}ri}},
  \bibinfo{author}{\bibfnamefont{C.}~\bibnamefont{De~Dominicis}},
  \bibnamefont{and} \bibinfo{author}{\bibfnamefont{I.~R.}
  \bibnamefont{Pimentel}}, \bibinfo{journal}{Eur. Phys. J. B}
  \textbf{\bibinfo{volume}{25}}, \bibinfo{pages}{361} (\bibinfo{year}{2002}),
  \eprint{cond-mat/0202162}.

\bibitem[{\citenamefont{Aspelmeier and Moore}(2003)}]{As_Mr}
\bibinfo{author}{\bibfnamefont{T.}~\bibnamefont{Aspelmeier}} \bibnamefont{and}
  \bibinfo{author}{\bibfnamefont{M.}~\bibnamefont{Moore}},
  \bibinfo{journal}{Phys. Rev. Lett.} \textbf{\bibinfo{volume}{90}},
  \bibinfo{pages}{177201} (\bibinfo{year}{2003}).

\bibitem[{\citenamefont{M{\'e}zard et~al.}(1987)\citenamefont{M{\'e}zard,
  Parisi, and Virasoro}}]{MePaVi}
\bibinfo{author}{\bibfnamefont{M.}~\bibnamefont{M{\'e}zard}},
  \bibinfo{author}{\bibfnamefont{G.}~\bibnamefont{Parisi}}, \bibnamefont{and}
  \bibinfo{author}{\bibfnamefont{M.~A.} \bibnamefont{Virasoro}},
  \emph{\bibinfo{title}{Spin Glass Theory and Beyond}},
  vol.~\bibinfo{volume}{9} of \emph{\bibinfo{series}{Lecture Notes in Physics}}
  (\bibinfo{publisher}{World Scientific}, \bibinfo{address}{Singapore},
  \bibinfo{year}{1987}).

\bibitem[{\citenamefont{de~Almeida and Thouless}(1978)}]{AT}
\bibinfo{author}{\bibfnamefont{J.~R.~L.} \bibnamefont{de~Almeida}}
  \bibnamefont{and} \bibinfo{author}{\bibfnamefont{D.~J.}
  \bibnamefont{Thouless}}, \bibinfo{journal}{J. Phys. A}
  \textbf{\bibinfo{volume}{11}}, \bibinfo{pages}{983} (\bibinfo{year}{1978}).

\bibitem[{\citenamefont{Temesv\'ari}(2008)}]{AT2008}
\bibinfo{author}{\bibfnamefont{T.}~\bibnamefont{Temesv\'ari}},
  \bibinfo{journal}{\prb} \textbf{\bibinfo{volume}{78}},
  \bibinfo{pages}{220401(R)} (\bibinfo{year}{2008}), \eprint{arXiv:0809.1839}.

\bibitem[{\citenamefont{Temesv\'ari}(2006)}]{droplet}
\bibinfo{author}{\bibfnamefont{T.}~\bibnamefont{Temesv\'ari}},
  \bibinfo{journal}{J. Phys. A} \textbf{\bibinfo{volume}{39}},
  \bibinfo{pages}{L61} (\bibinfo{year}{2006}), \eprint{cond-mat/0510209}.

\bibitem[{\citenamefont{Temesv\'ari}(2007)}]{nucl}
\bibinfo{author}{\bibfnamefont{T.}~\bibnamefont{Temesv\'ari}},
  \bibinfo{journal}{Nucl.\ Phys.\ B} \textbf{\bibinfo{volume}{772}},
  \bibinfo{pages}{340} (\bibinfo{year}{2007}), \eprint{arXiv:cond-mat/0612523}.

\bibitem[{\citenamefont{Parisi}(2004)}]{Parisi04}
\bibinfo{author}{\bibfnamefont{G.}~\bibnamefont{Parisi}},
  \bibinfo{journal}{Int. J. Mod. Phys. B} \textbf{\bibinfo{volume}{18}},
  \bibinfo{pages}{733} (\bibinfo{year}{2004}).

\bibitem[{\citenamefont{De~Dominicis et~al.}(1998)\citenamefont{De~Dominicis,
  Kondor, and Temesv\'ari}}]{beyond}
\bibinfo{author}{\bibfnamefont{C.}~\bibnamefont{De~Dominicis}},
  \bibinfo{author}{\bibfnamefont{I.}~\bibnamefont{Kondor}}, \bibnamefont{and}
  \bibinfo{author}{\bibfnamefont{T.}~\bibnamefont{Temesv\'ari}},
  \emph{\bibinfo{title}{Beyond the Sherrington-Kirkpatrick Model}}
  (\bibinfo{publisher}{World Scientific}, \bibinfo{year}{1998}),
  vol.~\bibinfo{volume}{12} of \emph{\bibinfo{series}{Series on Directions in
  Condensed Matter Physics}}, p. \bibinfo{pages}{119},
  \eprint{cond-mat/9705215}.

\bibitem[{\citenamefont{Pytte and Rudnick}(1979)}]{PytteRudnick79}
\bibinfo{author}{\bibfnamefont{E.}~\bibnamefont{Pytte}} \bibnamefont{and}
  \bibinfo{author}{\bibfnamefont{J.}~\bibnamefont{Rudnick}},
  \bibinfo{journal}{Phys. Rev. B} \textbf{\bibinfo{volume}{19}},
  \bibinfo{pages}{3603} (\bibinfo{year}{1979}).

\bibitem[{\citenamefont{Parisi}()}]{private_communication_Parisi1}
\bibinfo{author}{\bibfnamefont{G.}~\bibnamefont{Parisi}},
  \bibinfo{note}{private communication}.

\bibitem[{\citenamefont{Parisi and Rizzo}(2009{\natexlab{b}})}]{Parisi_Rizzo_3}
\bibinfo{author}{\bibfnamefont{G.}~\bibnamefont{Parisi}} \bibnamefont{and}
  \bibinfo{author}{\bibfnamefont{T.}~\bibnamefont{Rizzo}}
  (\bibinfo{year}{2009}{\natexlab{b}}), \eprint{arXiv:0910.4553}.

\bibitem[{\citenamefont{M\'ezard and Parisi}(1991)}]{MePa}
\bibinfo{author}{\bibfnamefont{M.}~\bibnamefont{M\'ezard}} \bibnamefont{and}
  \bibinfo{author}{\bibfnamefont{G.}~\bibnamefont{Parisi}},
  \bibinfo{journal}{J. Phys. I (France)} \textbf{\bibinfo{volume}{1}},
  \bibinfo{pages}{809} (\bibinfo{year}{1991}).

\bibitem[{\citenamefont{Parisi}(1980)}]{Parisi80_truncated}
\bibinfo{author}{\bibfnamefont{G.}~\bibnamefont{Parisi}}, \bibinfo{journal}{J.
  Phys. A} \textbf{\bibinfo{volume}{13}}, \bibinfo{pages}{1887}
  (\bibinfo{year}{1980}).

\end{thebibliography}

\end{document}